\documentclass[twocolumn,showpacs,preprintnumbers,superscriptaddress,prl,amsmath,amssymb]{revtex4}

\usepackage{graphicx}
\usepackage{dcolumn}
\usepackage{bm}

\begin{document}

\title{Dy-V magnetic interaction and local structure bias on the complex spin and orbital ordering in Dy$_{1-x}$Tb$_{x}$VO$_3$ (x\,=\,0 and 0.2)}

\author{J.-Q. Yan}

\affiliation{ Materials Science and Technology Division, Oak Ridge National Laboratory, Oak Ridge, TN 37831, USA}
\affiliation{Department of Materials Science and Engineering, University of Tennessee, Knoxville, TN 37996, USA}

\author{H. B. Cao}
\affiliation{ Quantum Condensed Matter Division, Oak Ridge National Laboratory, Oak Ridge, TN 37831, USA}

\author{M. A. McGuire}
\affiliation{ Materials Science and Technology Division, Oak Ridge National Laboratory, Oak Ridge, TN 37831, USA}

\author{Y. Ren}
\affiliation{ X-ray Science Division, Argonne National Laboratory, Argonne, IL 60439, USA}

\author{B. C. Sales}
\affiliation{ Materials Science and Technology Division, Oak Ridge National Laboratory, Oak Ridge, TN 37831, USA}

\author{D. G. Mandrus}
\affiliation{ Materials Science and Technology Division, Oak Ridge National Laboratory, Oak Ridge, TN 37831, USA}\affiliation{Department of Materials Science and Engineering, University of Tennessee, Knoxville, TN 37996, USA}
\date{\today}
\begin{abstract}
The spin and orbital ordering in Dy$_{1-x}$Tb$_{x}$VO$_3$ (x\,=\,0 and 0.2) was studied by measuring x-ray powder diffraction, magnetization, specific heat, and neutron single crystal diffraction. The results show that G-OO/C-AF and C-OO/G-AF phases coexist in Dy$_{0.8}$Tb$_{0.20}$VO$_3$ in the temperature range 2\,K$\sim$60\,K and the volume fraction of each phase is temperature and field dependent. The ordering of Dy moments at T*\,=\,12\,K induces a transition from G-OO/C-AF to a C-OO/G-AF phase. Magnetic fields suppress the long range order of Dy moments and thus the C-OO/G-AF phase below T*. The polarized moments induced at the Dy sublattice by external magnetic fields couple to the V 3d moments and this coupling favors the G-OO/C-AF state. Also discussed is the effect of the Dy-V magnetic interaction and local structure distortion on the spin and orbital ordering in Dy$_{1-x}$Tb$_{x}$VO$_3$.

\end{abstract}

\maketitle

\section{Introduction}

The orbital degree of freedom and its coupling with spin, charge and lattice in transition metal oxides have been interesting topics in condensed matter physics for many years.\cite{Tokura2000} The orbital physics of Jahn-Teller active  \emph{e$_g$}-electrons has been extensively studied in manganites where intriguing phenomena were observed due to strong electron correlations. In \emph{R}VO$_3$ (\emph{R} = rare earth and Y) perovskites, the octahedral-site V$^{3+}$ ions with the electronic configuration of \emph{t$^2$e$^0$} have only  $\pi$-bonding t$_{2g}$-electrons that are Jahn-Teller active. All \emph{R}VO$_3$ members show a series of spin and orbital ordering transitions below room temperature.\cite{Miyasaka2003} Thus \emph{R}VO$_3$ perovskites offer an ideal platform to study the orbital physics of $\pi$-bonding t$_{2g}$-electrons. A recent observation of ferroelectricity in DyVO$_3$ suggests that \emph{R}VO$_3$ is a new family of magnetic multiferroics with strong coupling between electric polarization and spin/orbital ordered states.\cite{Zhang2012}

Two different types of orbital ordering were reported in \emph{R}VO$_3$; in each the \emph{xy} orbital is assumed to be occupied by one electron whereas the other electron takes \emph{yz} and \emph{zx} orbitals alternatively along [100] and [010] axes in the \emph{ab} plane. The antiphase stacking of \emph{ab} planes in G-type orbital order (G-OO, Fig.\,1(a)) is in  contrast to the in-phase stacking in C-type orbital order (C-OO, Fig.\,1(b)).\cite{BlakePRB2002, BlakePRL2001, RenPRB2003} The corresponding magnetic order, which is consistent with Goodenough-Kanamori rules,\cite{GoodenoughBook}\cite{Kanamori} is G-type (G-AF, Fig.\,1(b)) for the C-OO state and C-type (C-AF, Fig.\,1(a)) for G-OO state, respectively. \emph{R}VO$_3$ members with \emph{R}\,=\,La,\,...,\,Tb have the G-OO/C-AF as the ground state. However,  \emph{R}VO$_3$ members with \emph{R}\,=\,Dy,\,...,\,Lu and Y show at low temperatures a phase switch from G-OO/C-AF to C-OO/G-AF while cooling. Obviously, DyVO$_3$ is located near the boundary, and the delicate competition between the two distinct spin/orbital ordered states can be disturbed by external stimuli such as magnetic field or pressure.\cite{RVO3HighPressure,ZhouPRL2007,Miyasaka2007PRL,Fujioka2010PRB}

The complex spin and orbital ordering in DyVO$_3$ has been extensively studied by measuring the magnetic properties, thermal transport, structure, dielectric response, and optical properties.\cite{RVO3HighPressure,Miyasaka2007PRL} It's believed that 4 transitions take place in DyVO$_3$: the G-type orbital order at T$_{OO}$\,=\,190\,K, the C-type magnetic order at T$_N$\,=\,113\,K, the transition from G-OO/C-AF to C-OO/G-AF at T$_{CG1}$\,=\,57(cooling)/64(heating)\,K, and a re-entrant transition to G-OO/C-AF at T$_{CG2}$\,=\,12(cooling)/22(heating)\,K. Strong hysteresis was found for the latter two transitions which are first order and accompanied by a structural transition between orthorhombic (Pbnm) for C-OO/G-AF and monoclinic (P2$_1$/b) for G-OO/C-AF. High pressure favors the C-OO/G-AF state which has a smaller volume.\cite{RVO3HighPressure,ZhouPRL2007} External magnetic fields applied along \emph{a}- or \emph{b}-axis favor the G-OO/C-AF state; in a magnetic field above 30\,kOe, only the G-OO/C-AF state is observed below T$_N$.\cite{Miyasaka2007PRL,Fujioka2010PRB} The picture that has emerged from several studies on \emph{R}VO$_3$ (\emph{R} = Dy, and Ho) suggests that both lattice distortions and Heisenberg magnetic exchange between the V 3d and \emph{R} 4f moments need to be considered in order to understand the field-induced phase transitions.\cite{Fujioka2010PRB}

A recent single crystal neutron diffraction study provided direct evidence for a noncollinear, weak ferromagnetic order of the Dy sublattice below 12\,K. The observed weak ferromagnetic order of the Dy sublattice suggests an effective coupling of the external magnetic fields and the Dy sublattice. However, in sharp contrast to previous reports\cite{RVO3HighPressure,Miyasaka2007PRL} where G-OO/C-AF is believed to be the ground state for V-sublattice, this neutron diffraction study suggests that the ground state is C-OO/G-AF for the V-sublattice.

Besides the above discrepancy about the spin and orbital ordered ground state in DyVO$_3$, how Dy$^{3+}$ long range magnetic order responds to external magnetic fields is still unknown. In addition, a study on how small perturbation of the Dy sublattice affects the delicate balance between different spin and orbital ordered states is missing. Motivated by the above questions, we studied the spin and orbital ordering in Dy$_{1-x}$Tb$_{x}$VO$_3$ (x\,=\,0 and 0.2). The partial substitution of Dy by Tb is expected to disturb the long range order of Dy moments, and introduce local lattice distortion by the size difference between 12-oxygen coordinated Dy$^{3+}$ and Tb$^{3+}$ ions. Our results demonstrate that (1) Below 12\,K, the long range magnetic order of Dy$^{3+}$ moments in DyVO$_3$ favors the C-OO/G-AF phase for the V-sublattice, which introduces a small fraction of C-OO/G-AF phase in the dominant G-OO/C-AF matrix; (2) external magnetic fields suppress the long range order of Dy$^{3+}$ moments thus destabilizing the C-OO/G-AF phase in the Dy ordered state; (3) the polarized Dy$^{3+}$ moments favor the G-OO/C-AF phase. In addition, we find that partial substitution of Dy by Tb induces phase coexistence of G-OO/C-AF and C-OO/G-AF in a wide temperature range below T$_{CG1}$; the fraction of each phase is temperature and field dependent. The effects of Dy-V magnetic interactions and local structural distortions on the spin and orbital ordering are discussed.

\section{Experimental Details}
DyVO$_3$ and Dy$_{0.8}$Tb$_{0.2}$VO$_3$ single crystals were grown with the floating zone technique. \cite{YanPRL2004} A DyVO$_3$ single crystal grown in a similar manner was investigated in a previous study.\cite{RVO3HighPressure} Elemental analysis of the crystals was performed using a Hitachi TM-3000 tabletop electron microscope equipped with a Bruker Quantax 70 energy dispersive X-ray (EDX) system. The analysis confirmed that the as-grown crystal has the nominal stoichiometry within experimental error. X-ray powder diffraction on pulverized single crystals was performed using a PANalytical X'Pert Pro MPD powder X-ray diffractometer with a copper target in the temperature range 16\,K\,$\leq$\,T\,$\leq$300\,K. Close inspection of the powder diffraction patterns at room temperature revealed no impurity phases. For Dy$_{0.8}$Tb$_{0.2}$VO$_3$, all observed reflections could be indexed using the orthorhombic space group (Pbnm)  with lattice parameters of \emph{a}\,=\,5.305(1)\,${\AA}$, \emph{b}\,=\,5.607(2)\,${\AA}$, and \emph{c}\,=\,7.599(2)\,${\AA}$. The high-energy high-resolution X-ray powder diffraction was performed at 11ID-C, Advanced Photon Source, Argonne National Laboratory. Single crystal neutron diffraction was measured at HB-3A four-circle diffractometer at the High Flux Isotope Reactor at the Oak Ridge National Laboratory. A neutron wavelength of 1.003 {\AA} was used with a bent perfect Si-331 monochromator.\cite{HB3A} Magnetic properties were measured with a Quantum Design (QD) Magnetic Properties Measurement System in the temperature interval 2.0 K $\leq$ T $\leq$ 300 K. The temperature dependent specific heat data were collected using a 9 Tesla QD Physical Properties Measurement System in the temperature range of 1.9 K $\leq$ T $\leq$ 300 K.

\begin{figure} \centering \includegraphics [width = 0.47\textwidth] {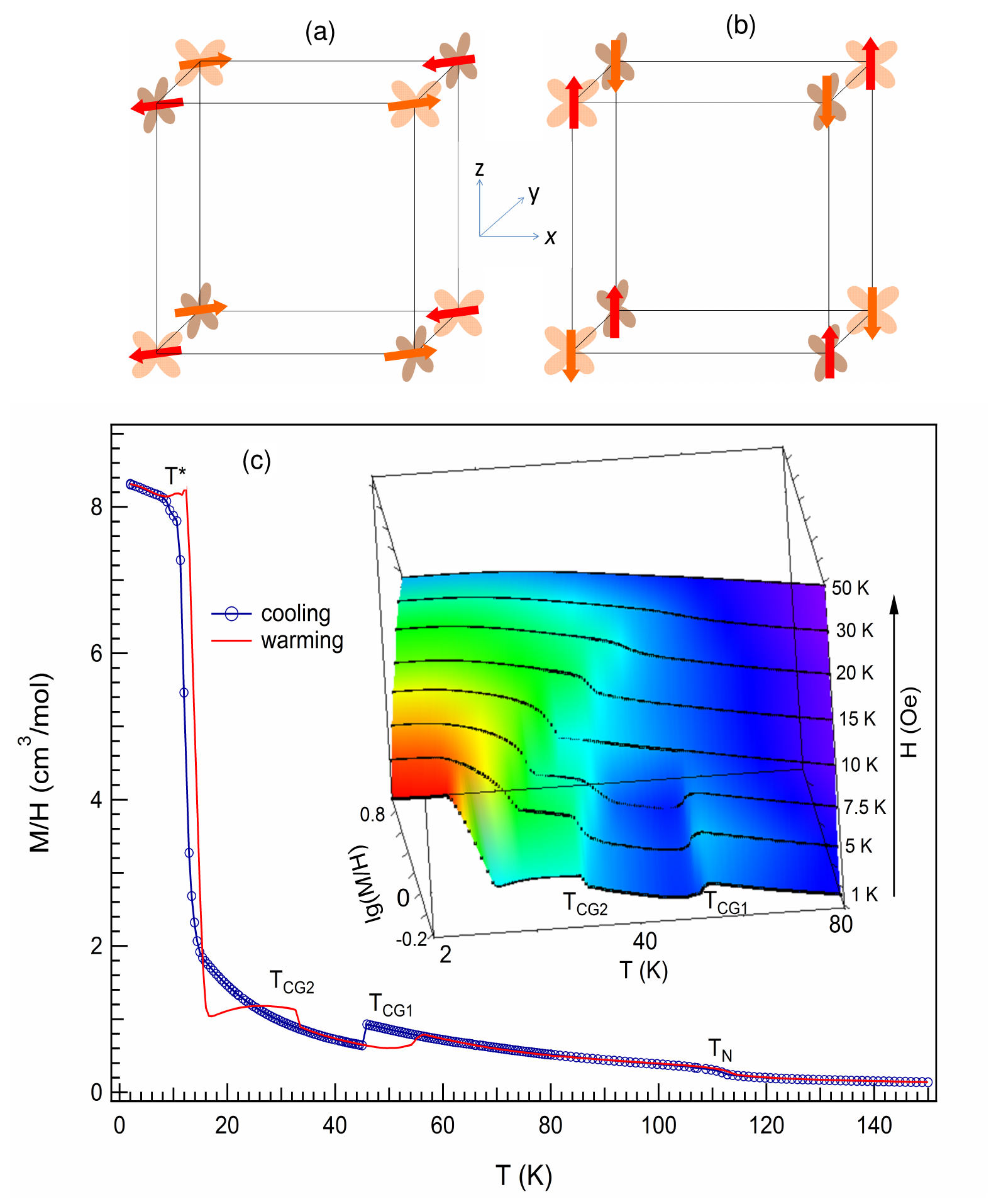}
\caption{(Color online) (a) Schematic diagram of the G-type orbital order (G-OO) with the C-type magnetic order (C-AF). (b)Schematic diagram of the C-type orbital order (C-OO) with the G-type magnetic order (G-AF). (c) The temperature dependence of M/H of Dy$_{0.8}$Tb$_{0.2}$VO$_3$ measured upon cooling and warming in a magnetic field of 1 kOe. The inset shows the magnetization curves measured in various magnetic fields upon warming.}
\label{Mag1K-1}
\end{figure}

\section{Experimental Results}
Figure\,\ref{Mag1K-1}(c) shows the temperature dependence of the M/H of Dy$_{0.8}$Tb$_{0.2}$VO$_3$ in an applied magnetic field of 1\,kOe measured both cooling and warming. The orientation of the sample is not well defined since Laue diffraction shows the crystal contains a few grains which prevents the study of anisotropic magnetic properties. The crystal was first cooled to 2\,K in zero magnetic field. At 2\,K, a magnetic field of 1\,kOe was applied and the magnetization was measured while warming to 300\,K. Then the magnetization data were collected while cooling from 300\,K to 2\,K. The rare earth moments dominate the paramagnetic state and no slope change was observed around T$_{OO}$ in the temperature dependence of H/M (not shown) for Dy$_{0.8}$Tb$_{0.2}$VO$_3$. For \emph{R}VO$_3$ members with nonmagnetic rare earth ions (\emph{R}\,=\,Y, Lu, Y$_{1-x}$La$_x$), a slope change in H/M is normally observed at T$_{OO}$.\cite{Ren2000PRB,Yan2005PRB,Yan2011PRB} As shown later in the temperature dependence of specific heat, T$_{OO}$ was determined to be 195\,K for Dy$_{0.8}$Tb$_{0.2}$VO$_3$. Below T$_{OO}$, the magnetization curve clearly shows multiple magnetic orderings. An increase of magnetization at 113\,K suggests the onset of long range magnetic order of the V-sublattice at T$_N$\,=\,113\,K. The increase of magnetization is due to a weak spin canting resulting from the Dzyaloshinsky-Moriya antisymmetric exchange interaction in the G-OO/C-AF phase. No hysteresis was was observed around T$_N$ which agrees with the second-order nature of this transition. Below T$_N$, two step-like anomalies at T$_{CG1}$=46\,K and T*=12\,K were observed with decreasing temperature; however, three could be well resolved at T*=14\,K, T$_{CG2}$=33\,K and T$_{CG1}$=56\,K with increasing temperature. The absence of T$_{CG2}$ in the cooling process is similar to that in DyVO$_3$. Compared to DyVO$_3$, Dy$_{0.8}$Tb$_{0.2}$VO$_3$ shows similar T*, T$_N$, and T$_{OO}$, but T$_{CG2}$ takes place at a higher temperature and T$_{CG1}$ occurs at a lower temperature. This suggests that the transition between C-OO/G-AF and G-OO/C-AF states at T$_{CG1}$ and T$_{CG2}$ is more sensitive to the rare earth site substitution. The step-like change and large hysteresis of magnetization at T$_{CG1}$ and T$_{CG2}$ agrees with the first-order nature of these two transitions. A small hysteresis at T* signals that the transition at T* is different from a simple regular second-order magnetic transition involving only the Dy sublattice.

\begin{figure} \centering \includegraphics [width = 0.47\textwidth] {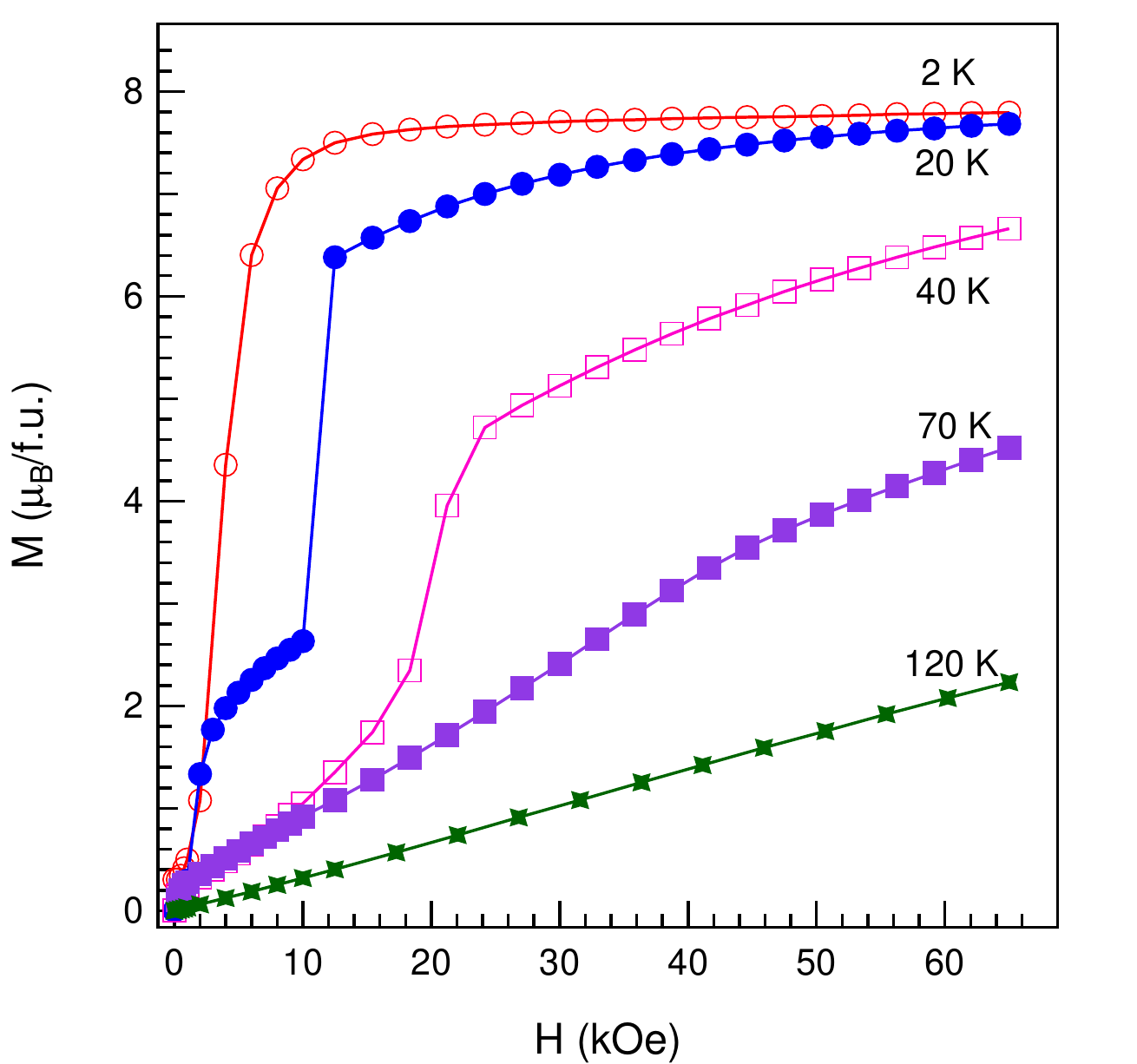}
\caption{(Color online) Field dependence of magnetization of Dy$_{0.8}$Tb$_{0.2}$VO$_3$ measured at indicated temperatures.}
\label{MH-1}
\end{figure}

The inset of Fig.\,\ref{Mag1K-1}(c) shows the M/H data of Dy$_{0.8}$Tb$_{0.2}$VO$_3$ measured upon warming in various magnetic fields. No field dependence of T$_N$ was observed in this study. The sharp change of M/H at T* broadens, decreases in magnitude, and shifts to higher temperatures with increasing magnetic fields. As this anomaly meets T$_{CG2}$ at H\,$\geq$\,10\,kOe, the step-like change at T$_{CG1}$ and T$_{CG2}$ disappears. It's worth mentioning that the field dependence for T$_{CG1}$ and T$_{CG2}$ is small before they disappear. This is in contrast to a large field dependence of T$_{CG2}$ in DyVO$_3$.\cite{Miyasaka2007PRL}

\begin{figure} \centering \includegraphics [width = 0.47\textwidth] {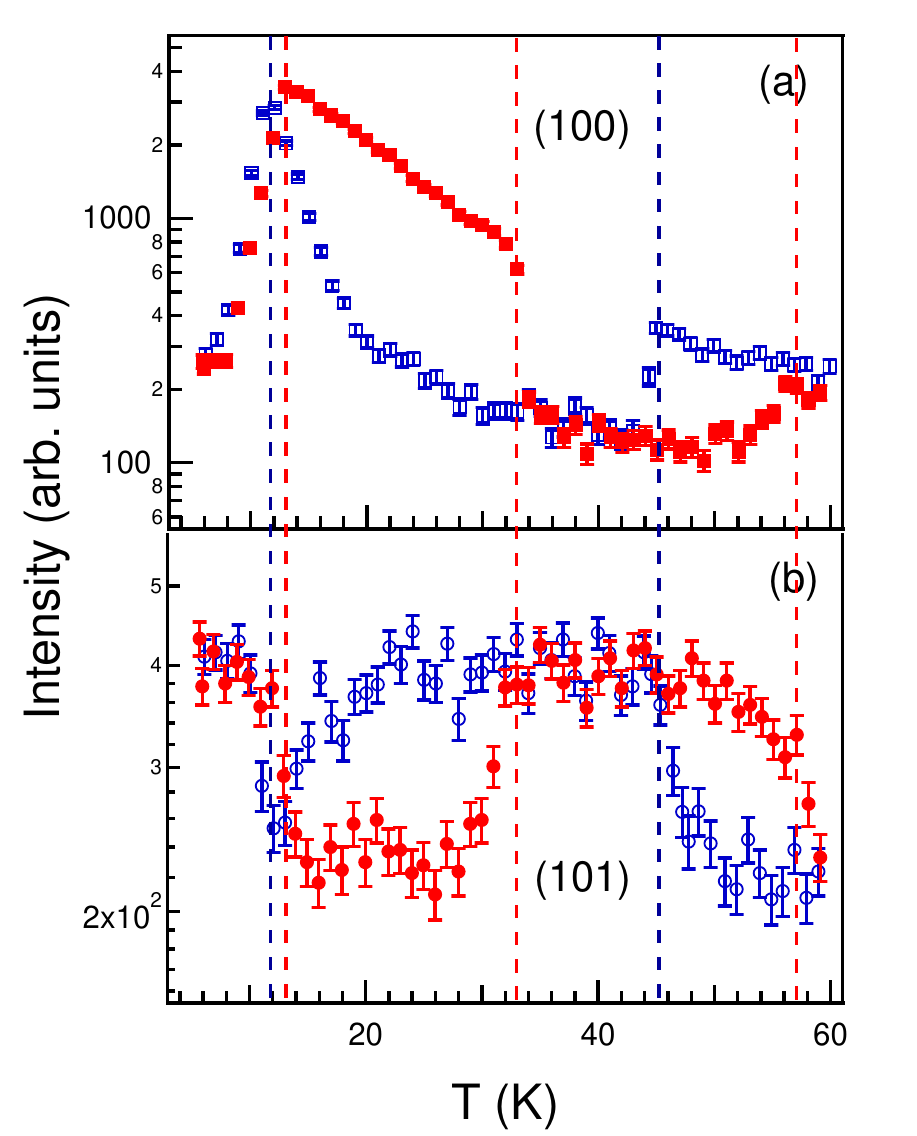}
\caption{(Color online) The temperature dependence of intensity of (100) and (101) magnetic reflections of Dy$_{0.8}$Tb$_{0.2}$VO$_3$ measured both warming (solid symbols) and cooling (open symbols). The vertical dashed lines highlight the temperatures where the intensity of (100) and (101) changes. The magnetic order of V sublattice in G-AF and C-AF ordered states could be distinguished by the (101) and (100) reflections, respectively.}
\label{VMag-1}
\end{figure}

Figure\,\ref{MH-1} shows the field dependence of magnetization measured at various temperatures. The jump in magnetization at low temperatures signals the occurrence of a metamagnetic transition. At 2\,K, a saturation moment of about 7.8 $\mu$$_B$ per formula unit was observed. Compared with the M(H) curves of DyVO$_3$,\cite{Miyasaka2007PRL} the data in Fig.\,\ref{MH-1} suggest that (1) the measurement is performed mainly along crystallographic b-axis, and (2) the metamagnetic transition is related to the rare earth 4f moments since a maximum magnetic moment of 2\,$\mu$$_B$ is expected for V$^{3+}$.

To confirm the magnetic orderings suggested by the M(T) data shown in Fig.\,1(c), single crystal neutron diffraction was performed in the temperature interval 4\,K$\leq$\,T\,$\leq$\,60\,K in zero magnetic field. From previous measurements,\cite{Reehuis201aPRB,YanPRL2007} the magnetic order of the V sublattice in G-AF and C-AF ordered states could be distinguished by the (101) and (100) (in orthorhombic notation) magnetic reflections, respectively. Figure\,\ref{VMag-1} shows the temperature dependence of intensity of (100) and (101) magnetic reflections measured upon both heating and cooling. As shown in Fig.\,\ref{VMag-1}(a), with increasing temperature, the intensity of (100) shows a sharp increase around 12\,K, then decreases slowly until 33\,K where a step-like drop takes place. At approximately 56\,K, the intensity of (100) shows another quick increase and reaches almost the same intensity as observed around 5\,K. When measured while cooling, the intensity of the (100) peak shows a step-like drop at 46\,K, then a gradual increase until $\sim$20\,K below which the intensity increases quickly to a maximum around 11\,K. Below 11\,K, the intensity quickly drops to a value similar to that observed at $\sim$60\,K. In contrast, the intensity of (101) magnetic reflection shows an opposite temperature dependence. With increasing temperature, the intensity shows a quick drop around 12\,K and a rapid increase around 32\,K, respectively. Above 30\,K it shows a sharp increase to a value similar to that below 10\,K, and then drops to be similar to that at 20\,K above 58\,K. During cooling, the intensity increases below 48\,K and shows a dip around 12\,K.

\begin{figure} \centering \includegraphics [width = 0.47\textwidth] {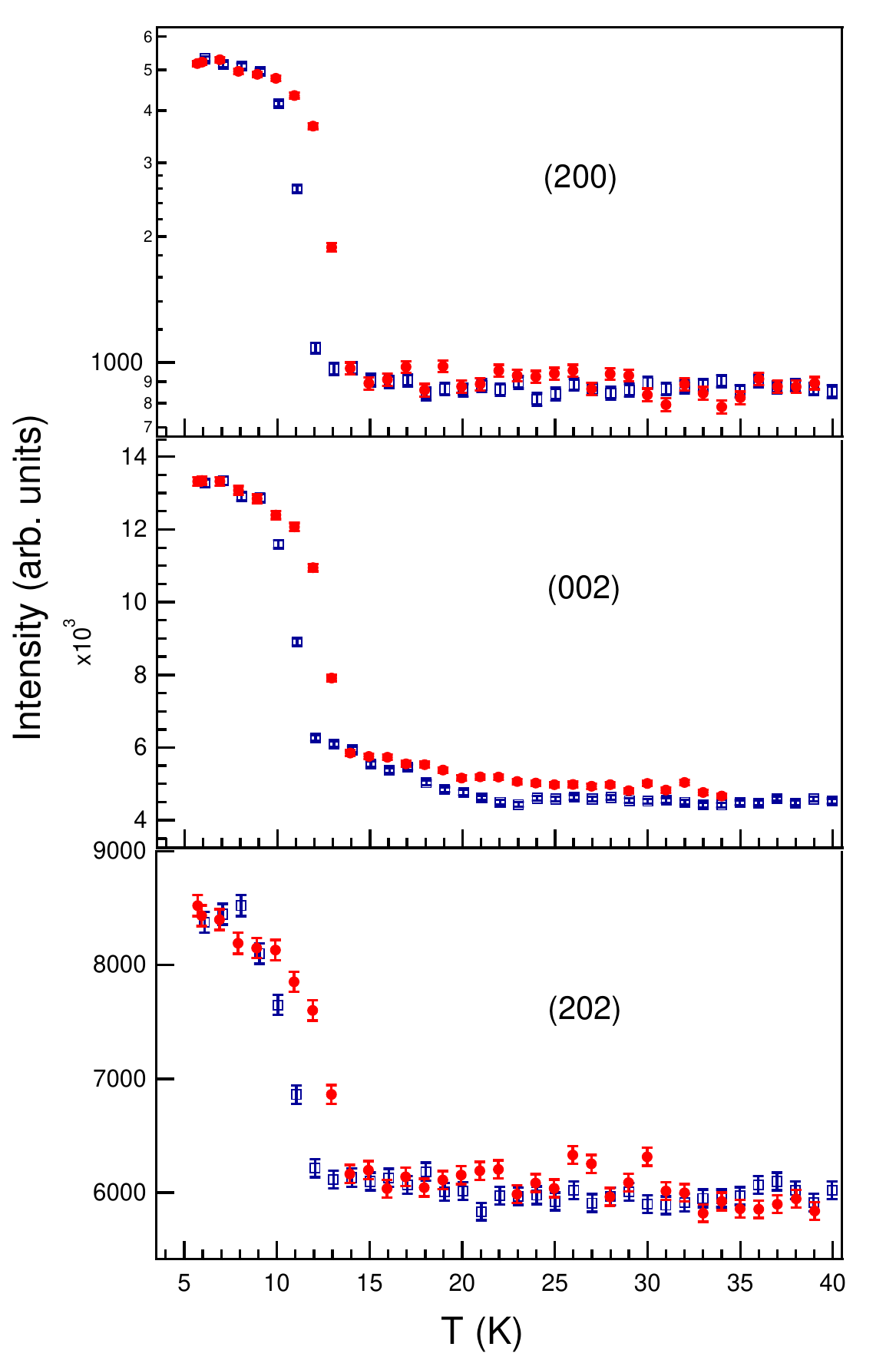}
\caption{(Color online) The temperature dependence of intensity of (200), (002) and (202) reflections of Dy$_{0.8}$Tb$_{0.2}$VO$_3$ measured upon both warming (solid symbols) and cooling (open symbols).}
\label{DyOrder-1}
\end{figure}

To confirm that Dy$^{3+}$ moments order at low temperatures, the temperature dependence of the (200), (002), and (202) reflections was monitored in the temperature range 4\,K$\leq$\,T\,$\leq$\,60\,K during both warming and cooling. These reflections are structurally allowed and an intensity change is expected at T*.\cite{Reehuis201aPRB} The results plotted in Fig.\,\ref{DyOrder-1} clearly show that Dy moments order below 12\,K and a small ($\sim$2\,K) hysteresis agrees with the observation in magnetization measurements shown in Fig.\ref{Mag1K-1}(c). The observed hysteresis is unusual since the Dy moment ordering is expected to be a continuous transition. This unusual feature is similar to that observed in DyVO$_3$ \cite{Reehuis201aPRB} and suggests that the long range magnetic order of Dy$^{3+}$ moments is accompanied with a transition between G-OO/C-AF and C-OO/G-AF states. Above T*, the peak intensity shows little difference when measured during cooling and warming. In contrast, in the temperature interval 12\,K$\leq$\,T\,$\leq$\,22\,K, (002) and (202) reflections of DyVO$_3$ are stronger in intensity when measured on warming than on cooling. The neutron single crystal diffraction study on DyVO$_3$ suggests a (C$_x$,F$_y$,-) and (F$_x$,C$_y$,-) type magnetic order of Dy$^{3+}$ moments below and above T*, respectively.\cite{Reehuis201aPRB} The results shown in Fig.\,\ref{DyOrder-1} suggests that partial substitution of Dy by Tb disturbs the ordering of Dy$^{3+}$ moments above T*.

\begin{figure} \centering \includegraphics [width = 0.47\textwidth] {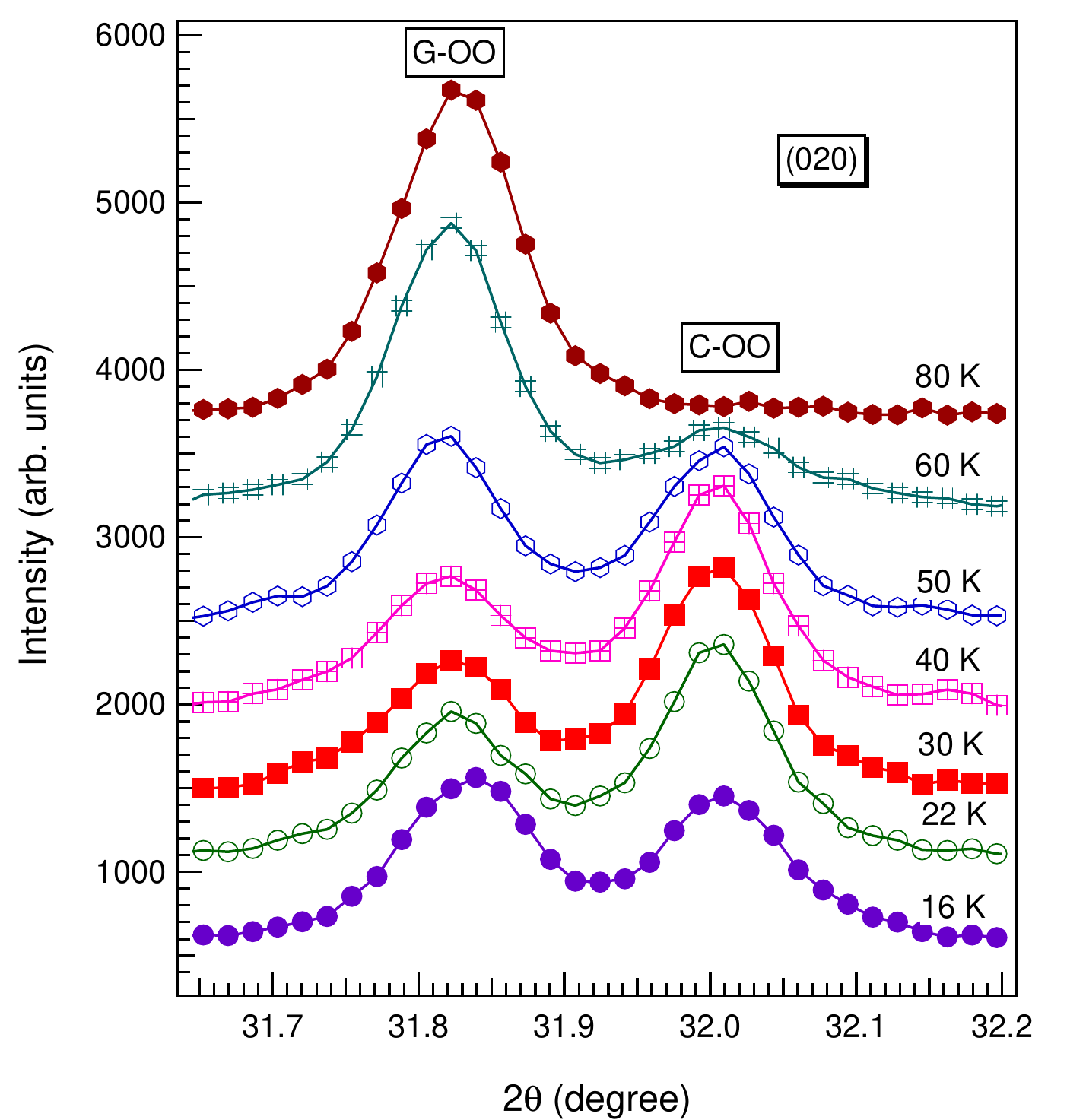}
\caption{(Color online) X-ray powder diffraction patterns of Dy$_{0.8}$Tb$_{0.2}$VO$_3$ measured using a  PANalytical X'Pert Pro MPD powder X-ray diffractometer with copper anode. The patterns were shifted vertically for clarity.}
\label{XRDMAM-1}
\end{figure}

\begin{figure} \centering \includegraphics [width = 0.47\textwidth] {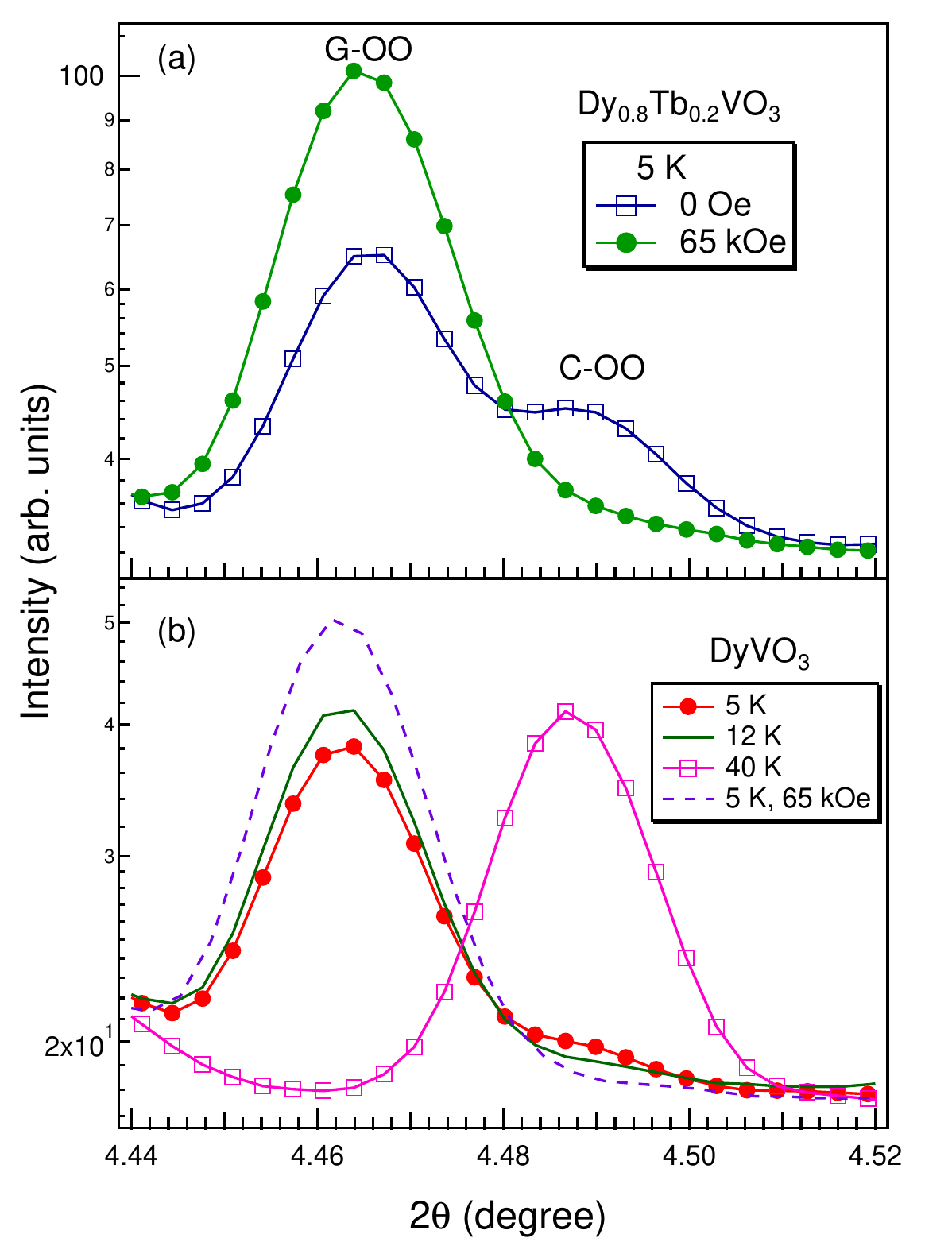}
\caption{(Color online) Evolution of (040)$_{orth}$ peak with temperature and magnetic field for (a) Dy$_{0.8}$Tb$_{0.2}$VO$_3$ and (b) DyVO$_3$. The high-energy high-resolution powder diffraction experiment was performed using a wavelength of $\lambda$=0.108${\AA}$.}
\label{Synchrotron-1}
\end{figure}

The temperature dependence of the intensity of the above studied magnetic reflections confirms in Dy$_{0.8}$Tb$_{0.2}$VO$_3$: (1) the long range magnetic order of Dy moments and the observation of C-OO/G-AF phase below T*; (2) on warming, the G-OO/C-AF state was observed in the temperature range 12\,K\,$\leq$\,T\,$\leq$\,32\,K, while the C-OO/G-AF state was favored for 32\,K\,$\leq$\,T\,$\leq$\,58\,K; (3)with decreasing temperature, the C-OO/G-AF state was observed in a wider temperature range below 42\,K. The above features are similar to those in DyVO$_3$.\cite{Reehuis201aPRB} However, the dip in the temperature dependence of the (101) peak around T* (see Fig.\ref{VMag-1}(b)) is absent in DyVO$_3$. The gradual change of the intensity of the (100) and (101) reflections below $\sim$20\,K suggests both the coexistence of G-OO/C-AF and C-OO/G-AF phases near T* and that the fraction of each phase is temperature dependent.

The transition from G-OO/C-AF to C-OO/G-AF is accompanied by a volume contraction. \cite{BlakePRB2002} To confirm the structural transitions and determine the phases in different temperature intervals, X-ray powder diffraction experiments were performed. Uniform powder pulverized from single crystals was first cooled down to 80\,K. Powder diffraction data were collected over a 2$\theta$ range of 10$^o$$\leq$2$\theta$$\leq$70$^o$ for selected temperatures from 80\,K to the base temperature at 16\,K. Figure\,\ref{XRDMAM-1} shows the temperature dependence of the (020)$_{orth}$ peak measured in the temperature range 16\,K$\leq$T$\leq$80\,K to highlight the phase evolution with temperature for Dy$_{0.8}$Tb$_{0.2}$VO$_3$. At 80\,K in the G-OO/C-AF state, a single peak at 2$\theta$=31.84\,degree was observed. At 60\,K, the peak at 2$\theta$=31.84\,degree drops in intensity and one new peak appears at 2$\theta$=32.01 degree signaling the appearance of the C-OO/G-AF phase. Below 60\,K, the coexistence of both G-OO/C-AF and C-OO/G-AF phases was well resolved with X-ray powder diffraction. The volume fraction ( roughly estimated from the peak intensity) of the C-OO/G-AF phase increases below 60\,K and then decreases below $\sim$30\,K. Our X-ray diffraction system was not able to measure below T*. To confirm the phase coexistence below T*, synchrotron X-ray powder diffraction was performed at 5\,K after cooling from room temperature (See Fig.\ref{Synchrotron-1}(a)). The double-peak feature agrees with the phase-coexistence scenario. The disappearance of the peak at 2$\theta$\,=\,4.49\,degree in a field of 65\,kOe signals that a large magnetic field can destroy the C-OO/G-AF phase. For comparison, Fig.\,\ref{Synchrotron-1}(b) shows the diffraction patterns of (040)$_{orth}$ peak for DyVO$_3$ collected at different temperatures. At 5\,K, a weak, broad peak at 2$\theta$=4.49\,degree signals the existence of small amount of C-OO/G-AF phase together with the majority G-OO/C-AF phase. The volume fraction of the C-OO/G-AF phase estimated from the peak intensity is much less than that in Dy$_{0.8}$Tb$_{0.2}$VO$_3$. With increasing temperature, this peak decreases in intensity and is barely observable at $\sim$12\,K; above T$_{CG2}$ it becomes dominant signaling the transition from the G-OO/C-AF state to the G-OO/C-AF state. As in Dy$_{0.8}$Tb$_{0.2}$VO$_3$, a large magnetic field wipes off the minority C-OO/G-AF phase at 5\,K. In contrast, the phase coexistence in DyVO$_3$ was only observed in the Dy ordered state below 12\,K. This clearly suggests that long range magnetic order of Dy moments favors the C-OO/G-AF state. We note that the G-OO/C-AF phase dominates below T* in DyVO$_3$. The small fraction of C-OO/G-AF phase is not stable under applied magnetic fields. The disappearance of the minor C-OO/G-AF phase below T* under applied magnetic fields in turn suggests that the long range order of Dy moments is suppressed by applied magnetic fields.

\begin{figure} \centering \includegraphics [width = 0.47\textwidth] {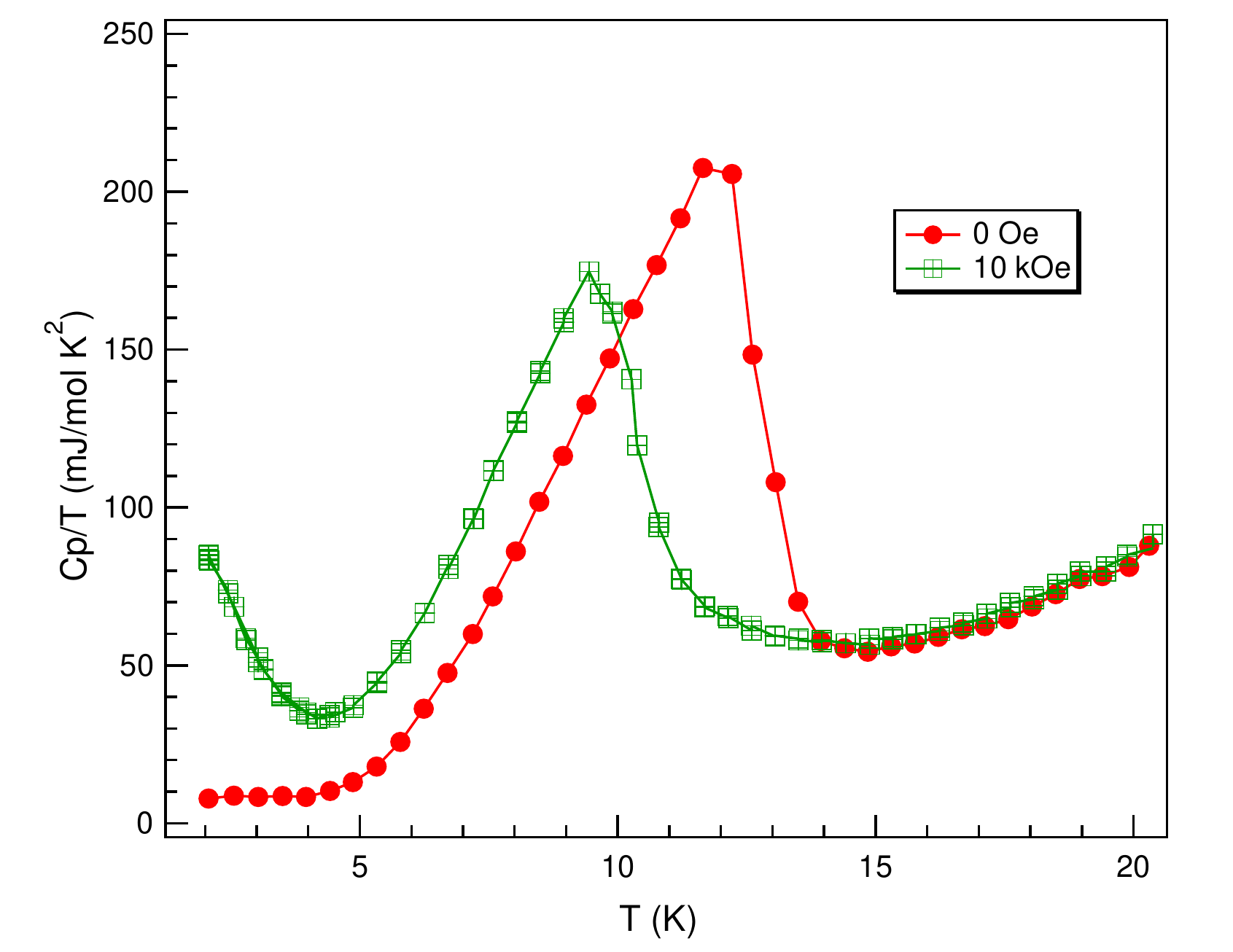}
\caption{(Color online) Magnetic field effect on the temperature dependence of specific heat of DyVO$_3$ near 12\,K. }
\label{CpDy-1}
\end{figure}

In order to further explore the magnetic field effect on the magnetic order of Dy moments, the specific heat of DyVO$_3$ was measured in 0\,Oe and 10\,kOe, respectively. As shown in Fig.\,\ref{CpDy-1} which plots Cp/T vs T, in zero field a lambda-type anomaly is well resolved at 12\,K (peak position) where Dy$^{3+}$ moments order. Once an external magnetic field of 10\,kOe is applied, the lambda-type anomaly shifts to about 9\,K and a Schottky-anomaly- like feature appears below 4\,K. The results clearly illustrate that external magnetic fields suppress the long range magnetic order of Dy moments. Measurements at even larger magnetic fields failed because the Apiezon N-grease failed to hold the sample on the stage due to the large torque that the sample experienced.

\begin{figure} \centering \includegraphics [width = 0.47\textwidth] {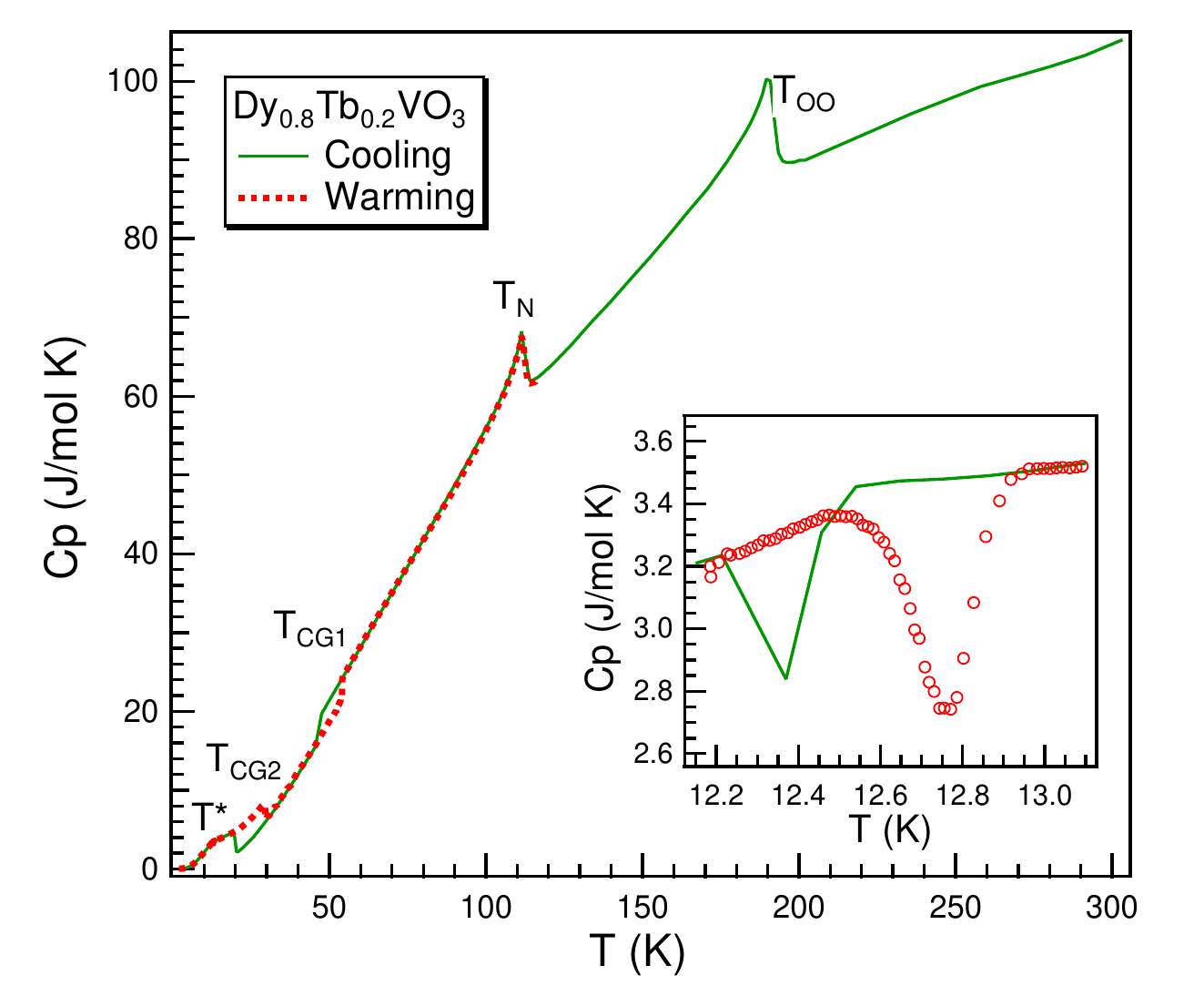}
\caption{(Color online) The temperature dependence of specific heat of Dy$_{0.8}$Tb$_{0.2}$VO$_3$. Inset highlights the details around 12\,K. }
\label{Cp-1}
\end{figure}

Figure\,\ref{Cp-1} shows the temperature dependence of specific heat of Dy$_{0.8}$Tb$_{0.2}$VO$_3$ measured in the temperature range 1.9\,K$\leq$\,T\,$\leq$\,300\,K. Two lambda-type anomalies could be well resolved at T$_{OO}$=195\,K and T$_N$=113\,K, respectively. No hysteresis was observed at T$_N$, which agrees with the magnetization measurement. When measured on warming, T$_{CG1}$ and T$_{CG2}$ determined from the step-like change agree with those obtained by magnetization and neutron measurements. With decreasing temperature, T$_{CG1}$ is lowered to 46\,K, and another step-like change was observed at 20\,K signaling the transition from C-OO/G-AF to G-OO/C-AF. We noticed that on cooling the intensity of the (100) peak shows a sharper increase while the intensity of the (101) peak starts to decrease below 20\,K, although no obvious anomaly was observed in the M/H curve in Fig.\ref{Mag1K-1}. These results suggest that T$_{CG2}$\,=\,20\,K on cooling. Around T*, Cp(T) shows a weak lambda-type anomaly. The inset of Fig. \ref{Cp-1} shows the details of Cp(T) around T*. Surprisingly, a dip on top of the broad lambda-type anomaly was observed in Cp(T) curve while both cooling and warming with a hysteresis of approximately 0.4\,K. This feature is unusual in the temperature dependence of specific heat. Considering that a lambda-type anomaly is expected for the continuous magnetic order of Dy moments and a step-like decrease of Cp(T) takes place at the transition from G-OO/C-AF to C-OO/G-AF, the unusual dip in Cp(T) curve suggests that a transition from G-OO/C-AF to C-OO/G-AF happens accompanying the magnetic ordering of Dy moments when cooling across T*. A detailed measurement of Cp(T) of DyVO$_3$ around T* failed to observe the dip-like feature as shown in the inset of Fig.\ref{Cp-1}. This is as expected because Cp(T) around T* is dominated by the well resolved lambda-type anomaly and also the volume fraction of the G-OO/C-AF phase that changes to C-OO/G-AF is small consistent with the X-ray powder diffraction data shown in Fig.\,\ref{Synchrotron-1}.

\section{Discussion}

In both Dy$_{0.8}$Tb$_{0.2}$VO$_3$ and DyVO$_3$, our study observed long range magnetic order of Dy moments and the coexistence of C-OO/G-AF and G-OO/C-AF phases below T*. Also, a series of spin and orbital ordering transitions take place above T* with increasing temperature at T$_{CG1}$, T$_{CG2}$, T$_{N}$, and T$_{OO}$, respectively. However, partial substitution of Dy by Tb induces the following differences between Dy$_{0.8}$Tb$_{0.2}$VO$_3$ and DyVO$_3$ that are noteworthy:

(1) In DyVO$_3$, the phase coexistence was observed only when Dy moments order below T*=12\,K. However,the coexistence of G-OO/C-AF and C-OO/G-AF phases was observed in a wide temperature interval below T$_{CG1}$=58\,K in Dy$_{0.8}$Tb$_{0.2}$VO$_3$. The volume fraction of each phase is temperature and field dependent.

(2) Tb substitution lowers T$_{CG1}$ but increases T$_{CG2}$ which means that the C-OO/G-AF phase is destabilized. T$_{CG2}$ could be resolved in Cp(T) data when measured during both heating and cooling. With decreasing temperatures, the anomalies in the intensities of (100), (101), and (002) reflections took place at about 20\,K where a step-like jump was observed in Cp(T) data.

(3) A large field dependence of T$_{CG2}$ was reported in DyVO$_3$.\cite{Miyasaka2007PRL} In contrast, T$_{CG2}$ of Dy$_{0.8}$Tb$_{0.2}$VO$_3$ shows little field dependence before it disappears when the external magnetic field is larger than 10\,kOe.

(4) A well-resolved lambda-type anomaly at T* in the Cp(T) curve of DyVO$_3$ was replaced by a weak anomaly in Dy$_{0.8}$Tb$_{0.2}$VO$_3$. The reduced entropy change across T* makes it possible to observe the specific heat change due to the phase transition from G-OO/C-AF to C-OO/G-AF driven by the ordering of Dy moments.

In \emph{R}VO$_3$ perovskites, the rare earth ions stay in the center of the dodecahedron formed by oxygen. The GdFeO$_3$-type distortion which involves the cooperative octahedral-site rotations is proportional to the ionic radius of the rare earth ions and affects the spin and orbital ordering of the V-sublattice.\cite{Mizokawa1996PRB,Mizokawa1999PRB} Partial substitution of Dy by Tb (1) increases the average ionic radius of the rare earth site, (2) introduces rare earth site variance due to the ionic radius difference between Dy$^{3+}$ and Tb$^{3+}$. The effect of size variance on the spin and orbital ordering in \emph{R}VO$_3$ perovskites has been studied in  Y$_{1-x}$La$_x$VO$_3$,\cite{YanPRL2007, Yan2011PRB} Y$_{1-x}$(La$_{0.23}$Lu$_{0.77}$)$_x$VO$_3$, \cite{YanPRL2007} Y$_{1-x}$Eu$_x$VO$_3$, \cite{Fujioka2010PRB} and Eu$_{1-x}$(La$_{0.254}$Y$_{0.746}$)$_x$VO$_3$.\cite{Fukuta2011PRB} Despite a different understanding of how quenched disorder affects the spin and orbital ordering, both Yan \emph{et al.}\cite{YanPRL2007, Yan2011PRB} and Fukuta \emph{et al.}\cite{Fukuta2011PRB} observed that the rare earth size variance suppresses T$_{OO}$ and T$_{N}$ but enhances T$_{CG}$. By contrast, the study of Y$_{1-x}$Eu$_x$VO$_3$ by Fujioka \emph{et al.}\cite{Fujioka2010PRB} showed that G-OO/C-AF phase is stabilized and C-OO/G-AF phase is destabilized with increasing \emph{x}. The substitution effect of Dy by Tb observed in this study is similar to that in Y$_{1-x}$Eu$_x$VO$_3$. Obviously, there are two distinct effects of quenched disorder on the spin and orbital ordering in \emph{R}VO$_3$ perovskites.

In all compositions mentioned above which show suppressed T$_{OO}$ and T$_{N}$ and enhanced T$_{CG}$ by quenched disorder, nonmagnetic La$^{3+}$ ions are present at the rare earth site. To rule out the possibility that this suppression is due to the nonmagnetic feature of La$^{3+}$ ions, we also studied the substitution effect of Y by Nd in YVO$_3$. Our preliminary study on the spin and orbital ordering in Y$_{1-x}$Nd$_x$VO$_3$ suggests that Nd substitution suppresses T$_{OO}$ and T$_{N}$, but enhances T$_{CG}$; 20\% Nd increases T$_{CG}$ to 85\,K as does 5\% La in Y$_{1-x}$La$_x$VO$_3$. These results strongly suggest that there is a critical local stress field above which the G-OO/C-AF phase is destabilized while the C-OO/G-AF phase is more stable. The partial substitution of Y by Eu in Y$_{1-x}$Eu$_x$VO$_3$ or Dy by Tb in Dy$_{1-x}$Tb$_{x}$VO$_3$ does increase the average ionic radius of rare earth site and induce size variance. However, the local stress field generated by the substitutional rare earth ions is below the critical value. Thus T$_{CG}$ is suppressed by the rare earth site disorder. The enhanced T$_{OO}$ and T$_{N}$ in Y$_{1-x}$Eu$_x$VO$_3$ suggests that T$_{OO}$ and T$_{N}$ mainly depend on the average structure, i.e, the increased V-O-V bond angle. Therefore, in considering the effect of quenched disorder on the spin and orbital ordering in \emph{R}VO$_3$ perovskites,  the magnitude of the local stress field generated by a single foreign atom is another factor in addition to the average ionic radius and size variance.

Despite suppressing T$_{CG1}$, the partial substitution of Dy by Tb favors the phase coexistence of C-OO/G-AF and G-OO/C-AF states below T$_{CG1}$\,=\,56\,K as revealed by X-ray powder diffraction patterns shown in Fig. \ref{XRDMAM-1} and \ref{Synchrotron-1}. This is in sharp contrast to DyVO$_3$ where the phase coexistence was observed only in the Dy$^{3+}$ ordered states below 12\,K. As discussed later, the long range magnetic order of Dy$^{3+}$ moments favors the C-OO/G-AF phase and introduces a phase transition from G-OO/C-AF to C-OO/G-AF. However, the disturbed Dy-V magnetic interaction by Tb substitution cannot explain (1) the phase coexistence in Dy$_{0.8}$Tb$_{0.2}$VO$_3$ in the temperature range below T$\leqq$T$_{CG2}$, or (2) more of the C-OO/G-AF phase in Dy$_{0.8}$Tb$_{0.2}$VO$_3$ below T* than for DyVO$_3$ since the long range order of Dy$^{3+}$ moments favor the C-OO/G-AF phase. We thus have to attribute the phase coexistence in a wide temperature range below T$_{CG1}$ to the local structure effect induced by Tb substitution. The phase coexistence further highlights the importance of local structure distortion in stabilizing the spin and orbital ordered states in DyVO$_3$ and agrees with the fact that the delicate balance between C-OO/G-AF and G-OO/C-AF states can be tuned with a small external stimulus. The evolution of the volume fraction of each phase with temperature suggests a temperature dependent change of the local structural distortion.

The long range order of Dy$^{3+}$ moments affects the spin and orbital ordering of the V-sublattice. In addition, strong external magnetic fields destroy the C-OO/G-AF phase. Two observations in this study suggest that the ordered Dy$^{3+}$ moments below T*=12\,K prefer the C-OO/G-AF state. First, synchrotron X-ray powder diffraction measurements reveal a small amount of the C-OO/G-AF phase in DyVO$_3$ below T*, which gradually disappears while warming above T*. Second, the dip-like feature in the temperature dependence of the specific heat of Dy$_{0.8}$Tb$_{0.2}$VO$_3$ could only be explained by a transition between the G-OO/C-AF and C-OO/G-AF phases riding on top of a weak lambda-type anomaly from the ordering of rare earth moments. The dip-like feature is only observable when the entropy change across T* is reduced by a proper amount of Tb substitution. The dominant G-OO/C-AF phase below T*, as revealed by X-ray powder diffraction results, suggests that this phase is still favored at the lowest temperatures. The field dependence of Cp(T) data from DyVO$_3$ demonstrates that applied magnetic fields suppress the long range order of Dy$^{3+}$ moments and thus the C-OO/G-AF phase. With increasing magnetic fields, Dy$^{3+}$ ordering is suppressed but there is a large magnetic polarization at each Dy site as well as a metamagnetic transition. The large polarized Dy$^{3+}$ moments interact with V$^{3+}$ 3d moments, which appears to favor the G-OO/C-AF phase.

\section{Conclusions}
We have investigated the effects of local structure distortions and Dy-V magnetic interactions on the spin and orbital ordering in Dy$_{1-x}$Tb$_{x}$VO$_3$ (x\,=\,0 and 0.2). Long range magnetic order of Dy$^{3+}$ moments  below 12\,K introduces a small fraction of the C-OO/G-AF phase in the G-OO/C-AF matrix. External magnetic fields suppress the long range order of Dy$^{3+}$ moments and induce a polarized moment on Dy$^{3+}$ sites which interacts with V$^{3+}$ 3d moments to stabilize the G-OO/C-AF phase. Local structure distortion induced by 20\% Tb substitution at the Dy site leads to the coexistence of C-OO/G-AF and G-OO/C-AF phases below T$_{CG1}$ and the volume fraction of each phase is field and temperature dependent. Compared with previous work on the effect of local structure effect on the spin and orbital ordering of Y$_{1-x}$La$_x$VO$_3$, Y$_{1-x}$(La$_{0.23}$Lu$_{0.77}$)$_x$VO$_3$, Y$_{1-x}$Eu$_x$VO$_3$, and Eu$_{1-x}$(La$_{0.254}$Y$_{0.746}$)$_x$VO$_3$, our study suggests that the magnitude of the local stress field around a single foreign atom is another tuning parameter in addition to the average ionic radius and macroscopic size variance.

We noticed that high quality DyVO$_3$ single crystals crack into pieces after cooling through the first-order transitions. However, with 20\% Dy substituted by Tb, the crystals can survive after more than 10 thermal cycles. Together with the fact that Tb substitution could be used to tune the volume fraction of each spin/orbital ordered phase, the doped compositions might provide a rich playground to explore field dependent ferroelectric properties.

\section{Acknowledgments}
JQY thanks H.D. Zhou and J.G. Cheng for helpful discussions and R. J. McQueeney for his support of crystal growth of DyVO$_3$ at Ames Laboratory. Work at ORNL was supported by the U.S. Department of Energy, Basic Energy Sciences, Materials Sciences and Engineering Division (JQY, MAM, BCS, and DGM) and the Scientific User Facilities Division (HBC). The use of beamline 11-ID-C at the Advanced Photon Source at Argonne National Laboratory was supported by the US Department of Energy, Office of Basic Energy Sciences under Contract No. DE-AC02-06CH11357.

\end{document}